%% file: Letter_scaling_v14.tex
\documentclass[final]{IEEEtran}

\usepackage{amsthm,amssymb,graphicx,multirow,amsmath,color,amsfonts}
\usepackage[update,prepend]{epstopdf}
\usepackage[noadjust]{cite}
\usepackage[latin1]{inputenc}
\usepackage{tikz}
\usetikzlibrary{arrows,calc}		
\usepackage{bbm} 
\usepackage{pdfpages}
\usepackage{tabulary}
\usepackage{multirow}
\usepackage{comment}

\usepackage{setspace}	

\def\chb#1{{\color{black} #1}}

\def\chbb#1{{\color{black} #1}}

\include{notation}

\allowdisplaybreaks 

\begin{document}
\graphicspath{{./Figures/}}

\title{
Equi-coverage Contours in  Cellular Networks}
\author{Mehrnaz Afshang, Chiranjib Saha, and Harpreet S. Dhillon
\thanks{The authors are with Wireless@VT, Department of ECE, Virgina Tech, Blacksburg, VA, USA. Email: \{mehrnaz, csaha,  hdhillon\}@vt.edu.}  \vspace{-.8em}
}


\maketitle
\vspace{-1.3ex}

\begin{abstract}
In this letter, we introduce a general cellular network model where i) users and BSs are distributed as two general point processes that may be coupled, ii) pathloss is assumed to follow a  multi-slope power-law  pathloss model, and iii) fading (power) is  assumed to be  independent across all wireless links.  For this setup, we first obtain a  set of contours  representing the same {\em meta distribution of \chb{$\sir$}}, which is the distribution of the conditional coverage probability  given the point process, for different  values of the parameters of the pathloss function and BS and user point processes. This general result is then specialized to 3GPP-inspired user and BS configurations obtained by combining Poisson point process (PPP) and Poisson cluster process (PCP).
\end{abstract}

\section{Introduction} \label{sec:intro}
Over the past decade, stochastic geometry has emerged as a powerful tool for the analysis  of cellular  networks. 
The basic principle of this approach is   to endow the locations of the users and base stations (BSs) with distributions (i.e., model them as {\em point processes}) and use the properties of the point processes to evaluate performance metrics such as  \chb{spatially averaged coverage probability, i.e., signal-to-interference-plus-noise  ratio ($\sinr$) distribution averaged  over point process}. In this letter, without going into any particular point process  analysis, we discover a useful relation between the point processes describing the locations  of the BSs and users, and power-law pathloss function for which the \chb{meta distribution of $\sir$~\cite{metadistribution}, i.e., distribution of the conditional coverage probability  given the point process,} of the cellular network remains the same. 

{\em Prior Art.} 
Most existing works focusing on the analysis of cellular networks using stochastic geometry model the locations of BSs and users as two independent   PPPs, and then evaluate \chb{spatially averaged coverage probability (in short coverage probability)} under the standard (single-slope) power-law  pathloss model, e.g. see~\cite{dhillon2012modeling}. One notable observation 
from \cite{dhillon2012modeling} is  that the \chb{coverage probability} ($\sinr$) does not depend on the BS density when the network is interference limited.
This  property  is termed as  {\em scale invariance} of  cellular networks  to  BS density. Follow up studies focusing on the effect of network scaling on the performance have taken two directions by relaxing one  of  the two assumptions: (i) using multi-slope power-law pathloss model\footnote{In multislope pathloss model,  path-loss  is a piece-wise power-law function with different exponents in different regions.} instead of single-slope path-loss~{\cite{MultiSlopeZhang2015,ding2016performance}}, and (ii)  using other point processes  to capture the spatial interactions of users and BSs~\cite{SahAfshDhiUnifiedHetNet2017}, which is completely ignored when their locations are modeled as PPPs. In both directions, \chb{coverage probability} departs significantly from the one observed in the baseline PPP-based model with single-slope pathloss. For instance, in a PPP-based network model, \chb{coverage probability} decreases beyond some  BS density if multi-slope pathloss is employed~\cite{MultiSlopeZhang2015}. For the other direction, consider a user-centric deployment scenario where small cell BSs (SBSs) are located at the center  of user hotspots (clusters). This can  be modeled by assuming that the users  follow a PCP and SBSs are  at the cluster centers  of the user PCP~\cite{SahAfshDhiUnifiedHetNet2017}. For this spatial setup, \chb{coverage probability} increases as the size of the user clusters (defined in terms of cluster radius or cluster variance of a PCP) decreases~\cite{SahAfshDhiUnifiedHetNet2017}.
 One can construct a more comprehensive analytical model  for cellular network by unifying these two aforementioned directions. For this unified model, in this letter, we demonstrate the existence of a simple scaling law involving the parameters of the  BS and user point processes and the boundaries of the pathloss model that  maintains  constant \chb{meta distribution of $\sir$}. \chbb{For such network models, this result provides  useful insights into the meta distribution of $\sir$ without the need to characterize it explicitly, which is challenging even for much simpler spatial models.} Specific contributions are summarized next.

{\em Contributions.} In this letter, we introduce general cellular network model with multi-slope pathloss under minimal assumptions on fading statistics, user and BS point processes, and cell association policy. For this model, we show that if the locations of users and BSs and the boundaries of the pathloss model are scaled simultaneously in a particular way, the \chb{meta distribution of $\sir$} of the network remains the same. We apply this  result  to  a few 3GPP-inspired cellular network models obtained  by  the combination of PPP and PCP and observe that  the scaling law generates equi-coverage contours in the parameter space of the network consisting of the density of PPP, cluster size of PCP representing user and/or BS point processes, and the boundaries of the  pathloss model.

\section{System Model} \label{sec:SysMod}
\subsubsection*{BS and user locations}
The cellular network consists of  BSs and users whose locations are distributed as two stationary  point processes $\Phi_{\rm b}({\bf p}_{\rm  b})$ and $\Phi_{\rm u}({\bf p}_{\rm u})$, where ${\bf p}_{\rm b}$ and ${\bf p}_{\rm u}$ denote the parameter sets associated with $\Phi_{\rm b}$ and $\Phi_{\rm u}$, respectively. Contrary  to the typical assumption of  independence of these two point processes~\cite{elsawy2013stochastic}, there may exist some spatial coupling between user and BS locations. One instance  of  spatial coupling between $\Phi_{\rm b}$ and $\Phi_{\rm u}$ is 
when users in $\Phi_{\rm u}$ form spatial clusters (or hotspots) and BSs are deployed at a higher density at the locations of these  hotspots. We will formally define the notion of independence and coupling  in the next Section.

\subsubsection*{Propagation model}
 We assume that all BSs transmit at power $P$ and  fading gain  between a BS at ${\bf b}\in\Phi_{\rm b}$ and a user at ${\bf u}\in\Phi_{\rm u}$  is denoted by $H_{{\bf b},{\bf u}}$, where $\{H_{{\bf b},{\bf u}}\}$ is a sequence of independently and identically distributed (i.i.d.) random variables, independent of user and BS point processes, with CDF $F_{H_{{\bf b},{\bf u}}}(\cdot)$.  Similar to \cite{MultiSlopeZhang2015}, for pathloss, we consider a piece-wise power-law function parameterized by ${\bf R} = [R_{c_0}, R_{c_1},...,R_{c_n}]$ and ${\pmb \alpha}=[\alpha_1, ..., \alpha_n]$: 
\begin{align} \label{eq:: general path-loss}
l(z,{\bf R})=\begin{cases}
\eta_1 z^{-\alpha_1} & 0=R_{c_0}<z\leq R_{c_1},\\
\eta_2 z^{-\alpha_2} & R_{c_1}<z\leq R_{c_2},\\
\vdots\\
\eta_n z^{-\alpha_n} & R_{c_{n-1}}<z< R_{c_n}=\infty,
\end{cases}
\end{align}
 with $z=\|{\bf b}-{\bf u}\|$, $\frac{\eta_j}{\eta_{j-1}} = R_{c_{j-1}}^{\alpha_j - \alpha_{j-1}}; \: \forall j>1$, $\alpha_0=0$, and $\eta_1$ is a  constant  which is assumed to be one without loss of generality.  Note that ${\bf R}$ and ${ \pmb \alpha}$ are functions of carrier frequency and the physical environment (indoor/outdoor). {For instance, $R_{c_1}$ is  approximately $4 h_{\rm t} h_{\rm r} f_{\rm c}/c$ in two-ray model {(which coincides with  the pathloss model with two slopes)}, where $h_{\rm t}$ and $h_{\rm r}$ are transmitter and receiver antenna heights, and $f_{\rm c}$ is the carrier frequency.} Note that by  setting $R_{c_1}\to\infty$, $l(z,{\bf R})$ in \eqref{eq:: general path-loss} reduces to the  single-slope pathloss model, which is  typically used in the  point process-based analysis of cellular networks~\cite{elsawy2013stochastic}. In this letter, we will refer to the set  $\{{\bf p}_{\rm u},{\bf p}_{\rm b},{\bf R}\}$ as {\em parameter space}, which represents all possible configurations of the cellular network model considered here.
 
For this general setup, we perform analysis on a typical user of  $\Phi_{\rm u}$. Considering the network to be interference-limited, we ignore thermal noise. The signal-to-interference ratio ($\sir$) experienced by a typical user located at ${\bf u}\in \Phi_{\rm u}$ is
\begin{align}\label{eq::sir::defn}
\sir_{\Phi_{\rm u},\Phi_{\rm b},l(z,{\bf R})}({\bf u})= \frac{ P H_{{\bf b}^*{\bf u}}  l(\|{\bf b}^{*}-{\bf u}\|,{\bf R})}{\sum_{{\bf b}  \in  \Phi_{\rm b} \setminus {\bf b}^{*}} P H_{{\bf b},{\bf u}}  l({\|\bf b}-{\bf u}\|,{\bf R})},
\end{align}
where ${\bf b}^{*}\in\Phi_{\rm b}$ denotes the location of the BS serving the user at ${\bf u}$, which is selected using one of the two cell association policies discussed next. 
\subsubsection*{Cell Association}
{While there exists a wide set of cell association policies for which our results will hold, to be concrete and avoid missing any corner cases, we limit our discussions to: (i) $\max$-power based association, where the typical user connects to the BS providing maximum average received power, i.e., ${\bf b}^* = \arg\max_{{\bf b}\in\Phi_{\rm b}} P l({\|\bf b}-{\bf u}\|,{\bf R})$ \cite{jo2012heterogeneous} and (ii) $\max$-$\sir$ based association, where the typical user connects to the BS  providing maximum instantaneous $\sir$, i.e.,  ${\bf b}^{*} = \arg\max\limits_{{\bf b}\in \Phi_{\rm b}, H_{{\bf b},{\bf u}} \in \{H_{{\bf b},{\bf u}}\}}P  {H_{{\bf b},{\bf u}}} l({\|\bf b}-{\bf u}\|,{\bf R})$~\cite{dhillon2012modeling}.}

\chbb{The conditional coverage probability of this network is ${\tt P}_{\rm c}=\P(\sir_{\Phi_{\rm u},\Phi_{\rm b},l(z,{\bf R})}({\bf u}) \ge \beta | {\bf u} \in \Phi_{\rm u}, \Phi_{\rm b} )$, where $\beta$ is  threshold for successful transmission. The {\em meta distribution} of $\sir$,  which is CCDF of the conditional coverage probability given the point process, is mathematically defined as: $\bar{F}_{\pc}(\Phi_{\rm u},\Phi_{\rm b},l({z},{\bf R}))=\E_{\Phi_{\rm b}}[{\bf 1} \{\P(\sir_{\Phi_{\rm u},\Phi_{\rm b},l(z,{\bf R})}({\bf u}) \ge \beta | {\bf u} \in \Phi_{\rm u}, \Phi_{\rm b} ) \ge \epsilon \}]$, 
 where $\epsilon \in [0,1]$~\cite{metadistribution}.}
 
 We now state our main problem statement: {\em when do two cellular networks  $(\Phi_{\rm u}({\bf p}_{\rm u}),\Phi_{\rm b}({\bf p}_{\rm b}),l({z},{\bf R}))$ and $(\Phi_{\rm u}({\bf p}_{\rm u}'),\Phi_{\rm b}({\bf p}_{\rm b}'),l(z, {\bf R}'))$ have the same \chb{meta distribution (or coverage)}, i.e., when is   $\bar{F}_{\pc}(\Phi_{\rm u}({\bf p}_{\rm u}),\Phi_{\rm b}({\bf p}_{\rm b}),l({z},{\bf R}))=\bar{F}_\pc(\Phi_{\rm u}({\bf p}_{\rm u}'),\Phi_{\rm b}({\bf p}_{\rm b}'),l({z},{\bf R}'))$?} In that case, $\{{\bf p}_{\rm u},{\bf p}_{\rm b},{\bf R}\}$ and $\{{\bf p}_{\rm u}',{\bf p}_{\rm b}',{\bf R}'\}$ will be said to lie on an equi-coverage contour in the parameter space.
\section {{Equi-coverage Contours}}\label{sec::equicov-system}
{\em Preliminaries}. We first recall Choquet's theorem~\cite{chiu2013stochastic} which states that a point process $\Phi$ in $\nbbR^2$ can be uniquely defined by its void probability, i.e., $\nbbP(\Phi(A) = 0)$, for any compact set $A\subset\nbbR^2$. $\Phi(A)$ denotes the associated counting measure of the point process, i.e., the number of points of $\Phi$ falling in $A$. 
  We now introduce the formal notion of  independence, equality, and scaling of  point processes as follows. 
  
 {\em Independence.} Two point processes $\Phi_{1}$ and $\Phi_{2}$ are independent iff $\nbbP(\Phi_{1}(A) = 0,\Phi_2(A) = 0)  =\nbbP(\Phi_{1}(A) = 0)\nbbP(\Phi_2(A) = 0)$ for any compact set $A\subset\nbbR^2$~\cite{chiu2013stochastic}.
 
 {\em Equality.} Two point processes $\Phi_{1}$ and $\Phi_{2}$ are equal in distribution iff their void probabilities are the same~\cite{chiu2013stochastic}, i.e.,
 \begin{align}
 \Phi_1 \stackrel{d}{=}\Phi_2 \Leftrightarrow  \nbbP(\Phi_{1}(A) = 0)=\nbbP(\Phi_2(A) = 0)
\end{align} 
for any compact set $A\subset\nbbR^2$~\cite{chiu2013stochastic}. Here `$\stackrel{d}{=}$' means equality in distribution.

{\em Point process-scaling}.  
Given a point process $\Phi=\{{\bf x}\}\subset\R^2$, the scaled process is denoted as $k\Phi = \{k{\bf x} : {\bf x} \in \Phi\}$ where $k\neq 0$ is a scalar. We now introduce the main result on the existence  of equi-coverage networks.
\begin{thm} 
\label{thm:: equality in coverage general}
Considering two stationary user and BS point processes $\Phi_{\rm u}$ and $\Phi_{\rm b}$, respectively,
\begin{equation}
\bar{F}_{\pc}(\Phi_{\rm u},\Phi_{\rm b},l(z,{\bf R})) = \bar{F}_{\pc}(k\Phi_{\rm u},k\Phi_{\rm b},l(z,{k} {\bf R})).\label{eq::coverage::thm}
\end{equation}
\end{thm}
\begin{IEEEproof} 
Let us denote  realizations of   $\Phi_{\rm b}$ and $\Phi_{\rm u}$  with $\phi_{\rm b}$ and $\phi_{\rm u}$, respectively.  Given $\phi_{\rm b}$ and $\phi_{\rm u}$, the  $\sir$ at a user location ${\bf u} \in \phi_{\rm u}$  is
\begin{align*}
&{\sir_{\phi_{\rm u},\phi_{\rm b},l({z},{\bf R})}}({\bf u}) \stackrel{}{=} \frac{  h_{{\bf b}^*, {\bf u}}  l(\|{\bf b}^{*}-{\bf u}\|,{\bf R})}{\sum_{{\bf b}  \in  \phi_{\rm b}\setminus {\bf b}^{*}}  h_{{\bf b}, {\bf u}}  l(\|{\bf b}-{\bf u}\|,{\bf R})}\\
&= \frac{  \sum_{i=1}^n{\bf 1}(R_{c_{i-1}}<\|{\bf b}^{*}-{\bf u}\| \le R_{c_{i}})  h_{{\bf b}^*, {\bf u}} \eta_i {\|{\bf b}^{*}-{\bf u}\|}^{-\alpha_i} }{\sum_{j=1}^n \sum\limits_{\substack{{\bf b}  \in  \phi_{\rm b}\setminus {\bf b}^{*}\\  R_{c_{j-1}} <\|{\bf b}-{\bf u}\|\le R_{c_j}}}h_{{\bf b}, {\bf u}} \eta_j   \|{\bf b}-{\bf u}\|^{-\alpha_j}}\\
&=  \sum\limits_{i=1}^n {\bf 1}(R_{c_{i-1}}<\|{\bf b}^{*}-{\bf u}\| \le R_{c_{i}}) \times \\
& \frac{   h_{{\bf b}^*, {\bf u}} \prod\limits_{\chb{\ell=2}}^{i} R_{c_{\ell-1}}^{\alpha_\ell- \alpha_{\ell-1}} {\|{\bf b}^{*}-{\bf u}\|}^{-\alpha_i} }{\sum_{j=1}^n\sum\limits_{\substack{{\bf b}  \in  \phi_{\rm b}\setminus {\bf b}^{*}\\  R_{c_{j-1}} <\|{\bf b}-{\bf u}\|\le R_{c_j}}} h_{{\bf b}, {\bf u}} \prod\limits_{\chb{\ell=2}}^{j}  R_{c_{\ell-1}}^{\alpha_\ell- \alpha_{\ell-1}}  \|{\bf b}-{\bf u}\|^{-\alpha_j}},
\end{align*}
where  $h_{{\bf b}^*, {\bf u}} $ and $h_{{\bf b}, {\bf u}} $ are realizations of $H_{{\bf b}^*, {\bf u}} $ and $H_{{\bf b}, {\bf u}} $.
Let us scale BS and user point process realizations by $k\phi_{\rm b} = \{k{\bf b}:{\bf b}\in\phi_{\rm b}\}$ and $k\phi_{\rm u} = \{k{\bf u}:{\bf u}\in\phi_{\rm u}\}$. \chb{The location of  the serving BS becomes $k {\bf b}^*$ for both max-$\sir$ and max-power cell association policies.} Now we can write
\begin{align*}
&\sir_{k\phi_{\rm u},k\phi_{\rm b},l({z},{\bf R})}(k{\bf u})= \sum\limits_{i=1}^n {\bf 1}(R_{c_{i-1}}<k \|{\bf b}^{*}-{\bf u}\| \le R_{c_{i}}) \times \\
& \frac{  h_{{\bf b}^*, {\bf u}} \prod\limits_{\chb{\ell=2}}^{i} R_{c_{\ell-1}}^{\alpha_\ell- \alpha_{\ell-1}} {(k \|{\bf b}^{*}-{\bf u}\|})^{-\alpha_i} }{\sum_{j=1}^n\sum\limits_{\substack{{\bf b}  \in  \phi_{\rm b}\setminus {\bf b}^{*}\\  R_{c_{j-1}} < k \|{\bf b}-{\bf u}\|\le R_{c_j}}}  h_{{\bf b},{\bf u}} \prod\limits_{\chb{\ell=2}}^{j} R_{c_{\ell-1}}^{\alpha_\ell- \alpha_{\ell-1}}  (k \|{\bf b}-{\bf u}\|)^{-\alpha_j}}\\
&\stackrel{(a)}{=}  \sum\limits_{i=1}^n {\bf 1}(R_{c_{i-1}}/k< \|{\bf b}^{*}-{\bf u}\| \le R_{c_{i}}/k)   \times\\
& \frac{ h_{{\bf b}^*, {\bf u}} \prod\limits_{\chb{\ell=2}}^{i} \big(R_{c_{\ell-1}}/k\big)^{\alpha_\ell- \alpha_{\ell-1}} { \|{\bf b}^{*}-{\bf u}\|}^{-\alpha_i} }{\sum\limits_{j=1}^n \hspace{-0.3cm} \sum\limits_{\substack{{\bf b}  \in  \phi_{\rm b}\setminus {\bf b}^{*}\\  R_{c_{j-1}}/k <  \|{\bf b}-{\bf u}\|\le R_{c_j}/k}}  \hspace{-0.3cm}  h_{{\bf b},{\bf u}}  \prod\limits_{\chb{\ell=2}}^{j} \big(R_{c_{\ell-1}}/ k\big)^{\alpha_\ell- \alpha_{\ell-1}}   \|{\bf b}-{\bf u}\|^{-\alpha_j}}\\
&=\sir_{\phi_{\rm u},\phi_{\rm b},l({z},k^{-1}{\bf R})}({\bf u}).
\end{align*}
From step (a), one can infer that scaling BS and user point processes with $k$ is equivalent to  scaling piece-wise function parameter set ${\bf R} = [R_{c_0}, R_{c_1},\dots,R_{c_n}]$ with $k^{-1}$. Therefore, if we simultaneously scale BS and user point processes as well as piece-wise function parameter set ${\bf R}$ with $k$, these two effects cancel each other, and  the $\sir$ becomes independent of $k$. More precisely, $
\sir_{ \phi_{\rm u}, \phi_{\rm b},l(z,{\bf R})}({\bf u}) = \sir_{k\phi_{\rm u}, k\phi_{\rm b},l(z,k{\bf R})}( k{\bf u}).$ Using this equivalence in  $\sir$s, we can establish following relation: \chb{$\E_{\Phi_{\rm b} }[{\bf 1}\{\P(\sir_{\Phi_{\rm u},\Phi_{\rm b},l(z,{\bf R})}({\bf u}) \ge \beta | {\bf u} \in \Phi_{\rm u}, \Phi_{\rm b}) > \epsilon\}]=
\E_{k\Phi_{\rm b} }[{\bf 1}\{\P(\sir_{k\Phi_{\rm u},k\Phi_{\rm b},l(z,{\bf R})}(k{\bf u}) \ge \beta | k {\bf u} \in k\Phi_{\rm u},  k\Phi_{\rm b})> \epsilon\}],$} which completes the proof. 
\end{IEEEproof} 

According to Theorem~\ref{thm:: equality in coverage general}, given $(\Phi_{\rm u},\Phi_{\rm u},l(z,{\bf R}))$, we can obtain  a sequence of  networks $\{(k\Phi_{\rm u},k\Phi_{\rm u},l(z,k{\bf R})),k\neq 0\}$ for which the \chb{meta distribution of $\sir$} will remain the same. This result points to the scale-invariance nature of \chb{meta distribution of $\sir$}. If a user and BS point process is simultaneously scaled with the same scale factor `$k$', \chb{$\bar{F}_{\pc}$} will remain the same if the boundaries of the pathloss function $l(z,{\bf R})$  are  also scaled by $k$. This scale-invariance property is independent of the particular user and BS point processes, their mutual coupling (or independence), and channel statistics (fading). We  conclude this discussion  by specializing the result of Theorem~\ref{thm:: equality in coverage general} for the case when  $\Phi_{\rm u}$ and $\Phi_{\rm b}$ are mutually independent and  stationary point processes.

%
\begin{cor} \label{cor: stationary and mutually independent}
  When  $\Phi_{\rm b}$ and $\Phi_{\rm u}$ are  stationary and mutually independent,  then it is sufficient to only scale BS point process and the boundaries of the pathloss model,  i.e., \chb{$\bar{F}_{\pc}(\Phi_{\rm u},\Phi_{\rm b},l(z,{\bf R})) 
=\bar{F}_{\pc}(\Phi_{\rm u},k\Phi_{\rm b},l(z,k{\bf R}))
.$}
\end{cor}
\begin{IEEEproof}
The proof simply follows from independence of $\Phi_{\rm u}$ and $\Phi_{\rm b}$.
\end{IEEEproof}
\begin{remark}
\label{Rem: equi-coverage contours}
\chbb{While~equi-coverage~contours~for~different association policies may be different,  the conditions for equi-coverage contours (appear in Theorem~\ref{thm:: equality in coverage general} and Corollary~\ref{cor: stationary and mutually independent})
 are the same for both max-$\sir$ and max-power cell association policies.}
\end{remark}
\subsubsection*{\bf {\em Equi-coverage contours in 3GPP cellular model}}
We now focus on the effect of scaling on  some point processes of interest for cellular network models (such as PPP and PCP). It has been shown in~\cite{SahAfshDhiUnifiedHetNet2017} that the combination of PPP and PCP can generate different representative models of cellular networks, which are consistent with the simulation models considered by 3GPP. We will first investigate  how scaling affects the parameter space of  PPP and PCP.  We denote  a homogeneous PPP by $\Psi(\lambda)$ since it is completely specified by  its density  $\lambda$ and  formally define a PCP as: 
\begin{ndef} A PCP, more precisely a Neyman-Scott point process is defined as 
$\Psi= \Phi_{\rm p}+ {\cal A}^{\bf x}, $
which is formed by a parent PPP $\{{\bf x}\}\equiv\Phi_{\rm p}(\lambda_{\rm p})$, where  offspring points ${\cal A}^{\bf x}$ are i.i.d. around each parent point located at ${\bf x} \in \Phi_{\rm p}$, and number of points per cluster $|{\cal A}^{\bf x}|$ is an independent  Poisson random variable with mean $\bar{m}$. The two popular special cases of PCPs are:
\begin{itemize}
\item Mat{\'e}rn Cluster Process (MCP) $\Psi(\lambda_{\rm p},\bar{m},r_{\rm d})$ : Points in ${\cal A}^{\bf x}$ are i.i.d with PDF
\begin{equation}\label{eq: PDF Matern}
f_{\bf S}({\bf s})= \frac{1}{\pi r_{\rm d}^2}; \quad \|{\bf s}\|\le r_{\rm d}.
\end{equation}

\item Thomas Cluster Process (TCP) $\Psi(\lambda_{\rm p},\bar{m},\sigma)$:  Points in ${\cal A}^{\bf x}$ are i.i.d with PDF
\begin{equation}\label{eq: PDF Thomas}
f_{\bf S}({\bf s})= \frac{1}{2 \pi \sigma^2}\exp \left(-\frac{\|{\bf s}\|^2}{2 \sigma^2}\right).
\end{equation}
\end{itemize}
\end{ndef} 
Note that, compared to a PPP $\Psi(\lambda)$,  TCP and MCP are specified by three parameters. While the first two parameters $\lambda_{\rm p}$ and $\bar{m}$ are common for all PCPs, the third parameter is specific to the point distribution around the cluster center. For MCP, $r_{\rm d}$ is the cluster radius and for TCP, $\sigma$ is the standard deviation of the locations of offspring points with respect to the cluster center. We now observe the effect of scaling on PPP and PCP in the following Lemma. 
  \begin{lemma} 
\label{label: lemma PPP-PCP k scaled}
When $\Psi$ is a homogeneous PPP with density $\lambda$ denoted by $\Psi(\lambda)$, then,
$k\Psi(\lambda) \stackrel{d}{=} \Psi(\lambda/ k^2).
$
When $\Psi$ is a PCP with parameters $(\lambda_{\rm p},\bar{m},\rho)$  denoted by $\Psi(\lambda_{\rm p},\bar{m},\rho)$, then,
$k\Psi(\lambda_{\rm p},\bar{m},\rho) \stackrel{d}{=} \Psi(\lambda_{\rm p}/k^2,\bar{m},\rho k),
$ where $\rho=\sigma$  and $\rho=r_{\rm d}$ for TCP and MCP, respectively.
\end{lemma} 
\begin{IEEEproof}See Appendix~\ref{app::PPP-PCP k scaled}.
\end{IEEEproof}

{\begin{remark}\label{rem::PPP-PCP scaling}
Lemma~\ref{label: lemma PPP-PCP k scaled} indicates that the scaled version of a PPP is another PPP with scaled density and the scaled version of an MCP (TCP) is an MCP (TCP) with scaled parent PPP density and cluster radius (standard deviation). 
\end{remark}}

Using the results on point process scaling discussed so far, we now find the equi-coverage networks for three cellular network models~\cite{SahAfshDhiUnifiedHetNet2017} defined as follows:

\textit{Model 1.} {\em User PPP-BS PPP} $(\Phi_{\rm u}(\lambda_{\rm u}),\Phi_{\rm b}(\lambda_{\rm b}),l(z,{\bf R}))$: $\Phi_{\rm u}$ and $\Phi_{\rm b}$ are independent PPPs of densities $\lambda_{\rm u}$ and $\lambda_{\rm b}$, respectively. 
 
\textit{Model~2.} {\em User~PCP-BS~PPP}~$(\Phi_{\rm u}(\lambda_{\rm p},\bar{m},\rho),\Phi_{\rm b}(\lambda_{\rm b}),l(z,{\bf R}))$: $\Phi_{\rm u}$ is a PCP ($\rho = r_{\rm d}$ for MCP and $\rho = \sigma$ for TCP) and $\Phi_{\rm b}$ is a PPP. {In this case, $\Phi_{\rm b}$ is assumed to be the parent PPP of $\Phi_{\rm u}$, and hence $\Phi_{\rm b}$ and $\Phi_{\rm u}$ are coupled, and $\lambda_{\rm p}=\lambda_{\rm b}$. 
This setup follows 3GPP's approach for modeling user hotspot and SBS locations where the SBSs are located at the center of the user hotspot~\cite{SahAfshDhiUnifiedHetNet2017}. 



%

\textit{Model 3.} {\em User~PCP-BS~PCP}~$(\Phi_{\rm u}(\lambda_{\rm p},\bar{m}_{\rm u},\rho),\Phi_{\rm b}(\lambda_{\rm p},\bar{m}_{\rm b},\rho),\\l(z,{\bf R}))$: 
$\Phi_{\rm u}$ and $\Phi_{\rm b}$ are both  PCPs, spatially coupled by the fact that both have the same parent PPP $\Phi_{\rm p}(\lambda_{\rm p})$. Also, both PCPs   $\Phi_{\rm u}$ and $\Phi_{\rm b}$ are conditionally  i.i.d. given the parent PPP. This setup is the representative of a scenario where multiple SBSs are deployed in a large user hotspot, e.g., airport, commercial complexes, to cover that area~\cite{SahAfshDhiUnifiedHetNet2017}.

%


For Models 1-3, the equi-coverage networks can be obtained by the following Proposition. 
\chb{\begin{prop} \label {Prop: equi-coverage Models1-3}The sequence of equi-coverage networks can be obtained as follows. For Model 1,
{\begin{align*}
{\bar{F}_\pc}(\Phi_{\rm u} (\lambda_{\rm u}),\Phi_{\rm b}(\lambda_{\rm b}),l(z,{\bf R})) = {\bar{F}_\pc}(\Phi_{\rm u} (\lambda_{\rm u}),\Phi_{\rm b}(
 \lambda_{\rm b}/k^2 ),l(z,k {\bf R})),
\end{align*} }
for Model~2,    ${\bar{F}_\pc} (\Phi_{\rm u} (\lambda_{\rm b}, \bar{m},   \rho),\Phi_{\rm b}(\lambda_{\rm b}),l({z},{\bf R}))=$
\begin{align*}
 {\bar{F}_\pc}(\Phi_{\rm u}(\lambda_{\rm b}/k^2, \bar{m},  k \rho  ),\Phi_{\rm b}( \lambda_{\rm b}/k^2 ),l({z},k {\bf R})),
\end{align*}
and for Model~3,   ${\bar{F}_\pc}(\Phi_{\rm u}(\lambda_{\rm p}, \bar{m},   \rho  ),\Phi_{\rm b}( \lambda_{\rm p} , \bar{m},  \rho ),l({z},{\bf R}))=$
\begin{align*}
 {\bar{F}_\pc}(\Phi_{\rm u}(\lambda_{\rm p}/k^2, \bar{m},  k \rho  ),\Phi_{\rm b}( \lambda_{\rm p}/k^2 , \bar{m},  k \rho ),l({z},k {\bf R})),
\end{align*}
where $\rho= \sigma_{\rm u}$ ($ r_{\rm d}$) for TCP (MCP), and $k\in\R^+$.
\end{prop}}
The analytical results presented in this letter are fairly self-explanatory and do not require a separate numerical treatment. For completeness, we plot   equi-coverage contours representing the same spatially averaged coverage probability $\bar{\tt P}_{\rm c}$ for Model~2 (TCP) with  single-slope ($n=1$) and dual-slope pathloss ($n=2$) in Figs. \ref{Fig:: equi-coverage line single pathloss} and \ref{Fig:: equi-coverage line dual pathloss}. Recall that spatially averaged coverage probability defined as:  $\E_{\Phi_{\rm b} }[\P(\sir_{\Phi_{\rm u},\Phi_{\rm b},l(z,{\bf R})}({\bf u}) \ge \beta | {\bf u} \in \Phi_{\rm u}, \Phi_{\rm b}) ]$, which is evaluated here  under max-$\sir$ cell association, when fading is assumed to be Rayleigh.


\begin{figure}
\minipage{0.23\textwidth}
  \includegraphics[width=\linewidth]{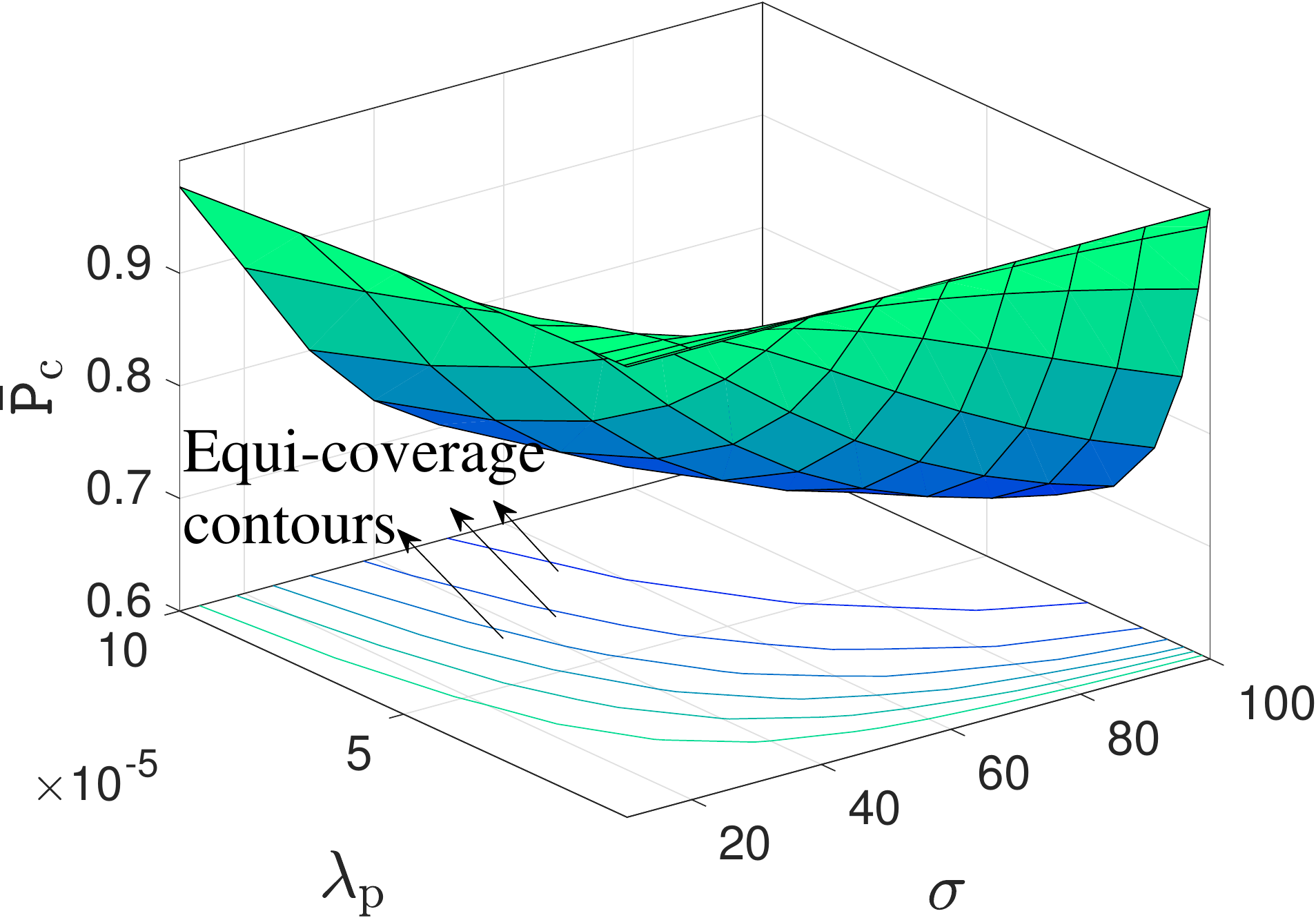}
  \caption{Equi-coverage contours for Model 2 with single-slope pathloss ($\alpha_1=4$)}\label{Fig:: equi-coverage line single pathloss}
\endminipage\hfill
\minipage{0.23\textwidth}
  \includegraphics[width=\linewidth]{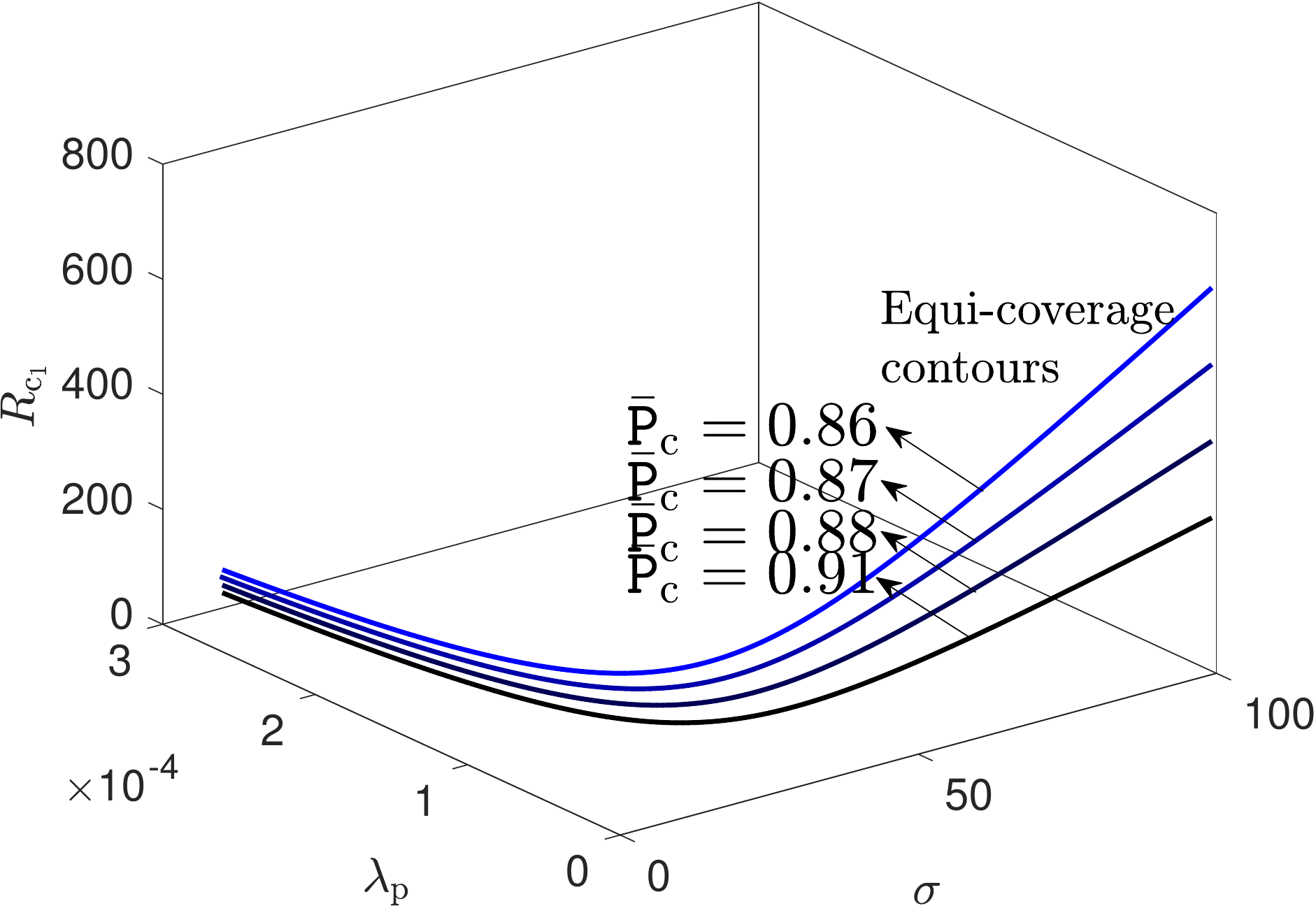}
  \caption{Equi-coverage contours for Model 2 with  dual-slope pathloss model. ($\alpha_1=3$,  $\alpha_2=4$) }\label{Fig:: equi-coverage line dual pathloss}
\endminipage
\end{figure}

%
%
%

%
%
%
%
%
%
%
%
%
%
%
\section{Concluding Remarks}
{In this letter, we introduced a general cellular network model with multi-slope pathloss  and  obtained equi-coverage contours in the parameter space of this model. While this comprehensive model captures a wide set of the morphologies of cellular networks considered in 3GPP simulations, 
its generality limits its analytical tractability (in terms of facilitating the exact analysis of key performance metrics, such as \chb{coverage probability}). Consequently, these equi-coverage contours are useful in gaining crisp insights into the trends of \chb{meta distribution of $\sir$} in the parameter space of the network. Further, these equi-coverage contours are a significant generalization of the density invariance of \chb{coverage probability} in PPP-based interference-limited  networks obtained under single-slope pathloss model. 
}



%

%
%

\appendix
\subsection{Proof of Lemma~\ref{label: lemma PPP-PCP k scaled}}
\label{app::PPP-PCP k scaled}
The equality  in distribution can be formally defined as \cite{chiu2013stochastic}:
\begin{align*}
\Phi_1 \stackrel{d}{=}\Phi_2  \Rightarrow \nbbP(\Phi_1(A) = 0) = \nbbP(\Phi_2(A) = 0),
\end{align*}
which states that  two point processes are equal  in distribution if the corresponding void probabilities are equal {for all compact sets $A \in \R^2$}.
For homogenous PPP, the void probability of the scaled point process, denoted by $k \Psi(\lambda)$, is
\begin{align*}
&\nbbP(|k\Psi(\lambda; A)|=0)= \E \Big[\prod_{{\bf z}\in \Psi} {\bf 1}(k {\bf y}\notin { A} ) \Big]\\
&\stackrel{(a)}{=}\exp \Big(-\lambda\int_{\R^2} \big(1-{\bf 1}(k {\bf y} \in A)\big) {\rm d} {\bf y}\Big)   \\
&\stackrel{(b)}{=} \exp \Big(-\lambda/ k^2 \int_{\R^2} \big(1-{\bf 1}({\bf y}' \in A)\big) {\rm d} {\bf y}'\Big)     
\end{align*}
where $(a)$ follows from PGFL of PPP \cite{haenggi2012stochastic}, and $(b)$ follows from ${\bf y}'=k {\bf y}$. From step $(b)$, we can infer that $k \Psi$ is a  homogenous PPP with density $\lambda/ k^2$. For scaled TCP, i.e.,  $k\Psi(\lambda_{\rm p}, \bar{m}, \sigma)$, the void probability can be expressed as:
\begin{align}\notag
&\P(|k\Psi(\lambda_{\rm p}, \bar{m}, \sigma ; A)|=0)=  \nbbE \bigg[\prod\limits_{{\bf x}\in\Phi_{{\rm p}}}\prod\limits_{{\bf y}\in{\bf x}+\ncalA^{\bf x}}{\bf 1}(k{\bf  y}\notin A )\bigg]\\\notag
&\stackrel{(a)}{=}\exp \bigg( -\lambda_{\rm p} \int\limits_{\nbbR^2}\bigg(1-\exp\bigg(-\bar{m}\bigg(\int\limits_{\R^2}  \big(1-{\bf 1}(k{\bf  y}\notin A)\big)  \\ &\times 
f_{\bf S}({\bf y}-{\bf x})   {\rm d}{\bf y}\bigg)\bigg)\bigg){\rm d}{\bf x}\bigg)\label{eq::void of PCP}\\\notag
& \stackrel{(b)}{=}\exp \bigg( -  \lambda_{\rm p}/k^2 \int\limits_{\nbbR^2}\bigg(1-\exp\bigg(-\bar{m}\bigg(\int\limits_{\R^2}  \big(1-{\bf 1}({\bf  y}'\notin A)\big)  \\ &\times 
\frac{1}{2 \pi (k \sigma)^2}\exp \left(-\frac{ \|{\bf y}'-{\bf x}'\|^2}{2 (k \sigma)^2}\right)   {\rm d}{\bf y}'\bigg)\bigg)\bigg)  {\rm d}{\bf x}'\bigg)\notag
\end{align}
where $(a)$ follows from PGFL of PCP \cite{ganti2009interference} and (b) from substituting \eqref{eq: PDF Thomas} in \eqref{eq::void of PCP} along with ${\bf x}' \to k {\bf x}$ and ${\bf y}' \to k {\bf y}$. Therefore, $k\Psi$ is  equal in distribution with a TCP of parameter set $(\lambda_{\rm p}/ k^2,{\bar m}, {\sigma k})$. Similarly one can write the void probability of  scaled MCP, denoted by $k \Psi$, as follows:
\begin{align*}
&\P(|k\Psi(\lambda_{\rm p}, \bar{m}, r_{\rm d};A)|=0)\stackrel{(c)}{=}\exp \bigg( - \lambda_{\rm p}/ k^2 \int\limits_{\nbbR^2}\bigg(1-\exp\bigg(\\
\times &-\bar{m} \bigg(\int\limits_{\R^2}   \big(1-{\bf 1}({\bf  y}'\notin A) \big)  \frac{{\bf 1}( \|{\bf y}'-{\bf x}'\|<r_{\rm d} k) }{\pi (r_{\rm d} k)^2}   {\rm d}{\bf y}'\bigg)\bigg)\bigg) {\rm d}{\bf x}'\bigg),
\end{align*}
where $(c)$ follows from substituting   \eqref{eq: PDF Matern} in \eqref{eq::void of PCP} along with ${\bf x}' \to k {\bf x}$ and ${\bf y}' \to k {\bf y}$. Thus scaled MCP, $k\Psi$, is equal in distribution with the MCP with parameters set $(\lambda_{\rm p}/ k^2,{\bar m}, {r_{\rm d} k})$.
\bibliographystyle{IEEEtran}

\bibliography{Letter_scaling_v14.bbl}


\end{document}

%% file: notation.tex
\def\nb0{{\mathbf{0}}}
\def\nb1{{\mathbf{1}}}

\def\ncalA{{\mathcal{A}}}

\def\nbbE{{\mathbb{E}}}

\def\nbbP{{\mathbb{P}}}

\def\nbbR{{\mathbb{R}}}

\newtheorem{lemma}{Lemma}
\newtheorem{thm}{Theorem}

\newtheorem{ndef}{Definition}

\newtheorem{prop}{Proposition}
\newtheorem{cor}{Corollary}

\newtheorem{remark}{Remark}

\def\E{\mathbb{E}}

\def\P{\mathbb{P}}
\def\pc{\mathtt{P_c}}

\def\R{\mathbb{R}}

\def\sinr{\mathtt{SINR}}			

\def\sir{\mathtt{SIR}}

\def\l{\ell}







%% file: Letter_scaling_v14.bbl
\begin{thebibliography}{10}
\providecommand{\url}[1]{#1}
\csname url@samestyle\endcsname
\providecommand{\newblock}{\relax}
\providecommand{\bibinfo}[2]{#2}
\providecommand{\BIBentrySTDinterwordspacing}{\spaceskip=0pt\relax}
\providecommand{\BIBentryALTinterwordstretchfactor}{4}
\providecommand{\BIBentryALTinterwordspacing}{\spaceskip=\fontdimen2\font plus
\BIBentryALTinterwordstretchfactor\fontdimen3\font minus
  \fontdimen4\font\relax}
\providecommand{\BIBforeignlanguage}[2]{{%
\expandafter\ifx\csname l@#1\endcsname\relax
\typeout{** WARNING: IEEEtran.bst: No hyphenation pattern has been}%
\typeout{** loaded for the language `#1'. Using the pattern for}%
\typeout{** the default language instead.}%
\else
\language=\csname l@#1\endcsname
\fi
#2}}
\providecommand{\BIBdecl}{\relax}
\BIBdecl

\bibitem{metadistribution}
M.~Haenggi, ``The meta distribution of the {SIR} in {P}oisson bipolar and
  cellular networks,'' \emph{IEEE Trans. on Wireless Commun.}, vol.~15, no.~4,
  pp. 2577--2589, Apr. 2016.

\bibitem{dhillon2012modeling}
{H. S. Dhillon}, R.~K. Ganti, F.~Baccelli, and J.~G. Andrews, ``Modeling and
  analysis of {$K$}-tier downlink heterogeneous cellular networks,'' \emph{IEEE
  Journal on Sel. Areas in Commun.}, vol.~30, no.~3, pp. 550--560, Apr. 2012.

\bibitem{MultiSlopeZhang2015}
X.~Zhang and J.~G. Andrews, ``Downlink cellular network analysis with
  multi-slope path loss models,'' \emph{IEEE Trans. on Commun.}, vol.~63,
  no.~5, pp. 1881--1894, May 2015.

\bibitem{ding2016performance}
M.~Ding, P.~Wang, D.~L{\'o}pez-P{\'e}rez, G.~Mao, and Z.~Lin, ``Performance
  impact of {LoS} and {NLoS} transmissions in dense cellular networks,''
  \emph{IEEE Trans. on Wireless Commun.}, vol.~15, no.~3, pp. 2365--2380, 2016.

\bibitem{SahAfshDhiUnifiedHetNet2017}
C.~Saha, M.~Afshang, and H.~S. Dhillon, ``{3GPP}-inspired {HetNet} model using
  {Poisson} cluster process: Sum-product functionals and downlink coverage,''
  \emph{IEEE Trans. on Commun.}, to appear.

\bibitem{elsawy2013stochastic}
H.~Elsawy, E.~Hossain, and M.~Haenggi, ``Stochastic geometry for modeling,
  analysis, and design of multi-tier and cognitive cellular wireless networks:
  A survey,'' \emph{IEEE Commun. Surveys and Tutorials}, vol.~15, no.~3, pp.
  996--1019, 3th quarter 2013.

\bibitem{jo2012heterogeneous}
H.-S. Jo, Y.~J. Sang, P.~Xia, and J.~G. Andrews, ``Heterogeneous cellular
  networks with flexible cell association: A comprehensive downlink {SINR}
  analysis,'' \emph{IEEE Trans. on Wireless Commun.}, vol.~11, no.~10, pp.
  3484--3495, Oct. 2012.

\bibitem{chiu2013stochastic}
S.~N. Chiu, D.~Stoyan, W.~S. Kendall, and J.~Mecke, \emph{Stochastic Geometry
  and its Applications}, 3rd~ed.\hskip 1em plus 0.5em minus 0.4em\relax New
  York: John Wiley and Sons, 2013.

\bibitem{haenggi2012stochastic}
M.~Haenggi, \emph{Stochastic Geometry for Wireless Networks}.\hskip 1em plus
  0.5em minus 0.4em\relax Cambridge University Press, 2012.

\bibitem{ganti2009interference}
R.~K. Ganti and M.~Haenggi, ``Interference and outage in clustered wireless ad
  hoc networks,'' \emph{IEEE Trans. on Info. Theory}, vol.~55, no.~9, pp.
  4067--4086, Sep. 2009.

\end{thebibliography}
